\documentclass[prd,twocolumn,amsmath,amssymb,floatfix,superscriptaddress,nofootinbib]{revtex4-1}
\pdfoutput=1
\usepackage{graphicx}
\usepackage{amssymb}
\usepackage{amsmath}
\usepackage{bm}
\usepackage{color}
\usepackage{natbib}

\DeclareMathAlphabet\mathbfcal{OMS}{cmsy}{b}{n}
\definecolor{darkgreen}{RGB}{50,150,0}
\definecolor{purple}{cmyk}{0.5,0.75,0,0}

\newcommand{\rd}{{\rm d}}

\newcommand{\Pa}{{\cal G}_\rho}
\newcommand{\Pb}{{\cal G}_v}
\newcommand{\Pc}{{\cal G}_p}
\newcommand{\Pd}{{\cal G}_\pi}

\newcommand{\Comment}[1]{{}}

\definecolor{ultramarine}{rgb}{0.07, 0.04, 0.56}
\definecolor{cadmiumgreen}{rgb}{0.0, 0.42, 0.24}
\definecolor{indigo(dye)}{rgb}{0.0, 0.25, 0.42}
\usepackage[linktocpage=true]{hyperref}
\hypersetup{
colorlinks=true,
citecolor=ultramarine,
linkcolor=cadmiumgreen,
urlcolor=indigo(dye),
pdfauthor={},
pdftitle={},
pdfsubject={}
}

\begin{document}

\title{Separate Universes beyond General Relativity}

\author{Wayne Hu}
\affiliation{Kavli Institute for Cosmological Physics, Department of Astronomy \& Astrophysics,  Enrico Fermi Institute, University of Chicago, Chicago, IL 60637}

\author{Austin Joyce}
\affiliation{Kavli Institute for Cosmological Physics, Department of Astronomy \& Astrophysics,  Enrico Fermi Institute, University of Chicago, Chicago, IL 60637}
\affiliation{Center for Theoretical Physics, Department of Physics, Columbia University, New York, NY 10027}

\begin{abstract}
\noindent
We establish purely geometric or metric-based criteria for the validity of the separate
universe ansatz, under which the evolution of small-scale observables in a long-wavelength perturbation is indistinguishable from a separate Friedmann--Robertson--Walker cosmology in their angle average.   In order to be able to identify the local volume expansion and curvature in a long-wavelength
perturbation with those of the separate universe,  we show that the lapse perturbation must be much smaller in amplitude than the  curvature  potential on a time slicing that comoves with the Einstein tensor.  
Interpreting the Einstein tensor as an effective stress energy tensor, the condition is that the effective stress energy comoves with  freely falling synchronous
observers who establish the local expansion, so that the local curvature is conserved. By matching the expansion history of these synchronous observers in cosmological simulations, 
one can establish and test consistency relations even in the nonlinear regime of modified
gravity theories.
\end{abstract}

\maketitle

\section{Introduction}

In the separate universe approach, the impact of a long-wavelength cosmological perturbation on 
local, short-wavelength observables is modeled as a change in the background cosmology.   This technique has proven  very useful both conceptually and as a practical tool for constructing consistency
relations between observables that hold nonlinearly.   Separate universe arguments provide insights into  relations
between $N$-point functions \cite{Maldacena:2002vr, Creminelli:2004yq}, the evolution of
isocurvature fluctuations in multifield models \cite{Sasaki:1995aw,Wands:2000dp,Grin:2011tf,He:2015msa}, shifts in  baryon acoustic oscillations \cite{SherwinZaldarriaga:12}, super sample power spectrum covariance \cite{TakadaHu:13,Lietal:14a},
 position dependent
power spectra \cite{Lietal:14b,Chiang:2014oga,Chiang:2015eza},
CMB lensing covariance \cite{Manzotti:2014wca}, and
dark matter halo bias 
\cite{Baldauf:2011bh,Baldauf:2015vio,Li:2015jsz,Lazeyras:2015lgp}.   

In all of these studies, general relativity is assumed from the outset and so the Einstein equations allow the validity of the separate universe  to be established through a mix of conditions on the
metric and the stress-energy of the matter such that a local observer will see a constant
comoving local curvature \cite{Bardeen:1980kt,Kodama:1985bj,Weinberg:2003sw,Dai:2015jaa}, which is the essential requirement for the construction to work.  In the following, we elucidate purely geometric
 conditions for the validity of the separate universe ansatz, with an eye toward extending these applications to non-Einsteinian gravity theories.  

Previous works 
on this question have focused on the infinite wavelength limit, where the perturbed metric
is matched exactly to a separate Friedmann--Robertson--Walker (FRW) cosmology 
\cite{Bertschinger:2006aw} or have tested the ansatz in the context of specific models
\cite{Song:2006jk,Song:2006ej,Gleyzes:2013ooa,Gleyzes:2014rba} and
 parameterizations  \cite{Hu:2007pj,Hu:2008zd,Hojjati:2011ix,Zhao:2008bn,Baker:2015bva}
of modified gravity.    Since the scale at which the separate
universe ansatz ceases to hold determines where  observational
violations of consistency relations occur, e.g.\ scale dependent halo bias and
squeezed $N$-point functions \cite{Hu:2016ssz,Chiang:2016vxa}, we seek to establish conditions under which the
separate universe ansatz approximately holds.  In doing so, we also clarify the role of 
anisotropic stress, real or effective, and shear in the expansion in these conditions.

The outline of the paper is as follows.  In \S \ref{sec:local} we consider the geometric
correspondence between perturbations in the Einstein tensor and the local expansion and
curvature.   We  establish the comoving lapse condition for the validity of the 
separate universe ansatz and relate it to conservation of curvature for synchronous and comoving local observers in \S \ref{sec:curvature}.   In \S \ref{sec:scales}, we discuss
the scales associated with validity of the separate universe ansatz and discuss
these results in \S \ref{sec:discussion}.

\section{Local Expansion and Curvature}
\label{sec:local}

In this section, we establish the relationship between metric perturbations and the local expansion and curvature,
utilizing the
geometric interpretation of the Einstein tensor and its covariant conservation.   These associations can be made in any gauge and
any metric theory of gravity. However absorbing the perturbations into a local FRW expansion requires
further restrictions that highlight special choices of time slicing for \S \ref{sec:curvature}.
We work in the mostly plus metric convention throughout.

\subsection{Background Geometry}

Given a background FRW spacetime with line element,
\begin{equation}
\rd s^2= a(\eta)^2 (-\rd\eta^2 + \gamma_{ij} \rd x^i \rd x^j),
\end{equation}
 where $\gamma_{ij}$ is a $3$-metric of constant comoving curvature $K$,
the Einstein tensor is
\begin{eqnarray}
\label{eqn:Gbackground}
G^0_{{\hphantom 0}0} &=& -3\left(  H^2 + \frac{K}{a^2} \right) = -3 H^2 - \frac{R^{(3)}}{2},\\
G^i_{{\hphantom i}j} -G^0_{{\hphantom 0}0}\frac{ \delta^i_{{\hphantom i}j} }{3}&=& -\frac{2}{a^2} \left( \frac{\ddot a}{a} - {a^2 H^2}   \right) \delta^i_{{\hphantom i}j}
= -\frac{2}{a} \frac{\rd^2 a}{\rd t^2} \delta^i_{{\hphantom i}j} , \nonumber
\end{eqnarray}
where overdots denote derivatives with respect to the ``0", conformal time  component, $\eta=\int\rd t/a$,
and $H \equiv \rd \ln a/\rd t$.  Here $R^{(3)}=6K/a^2$ is the 3-Ricci scalar and defines the physical curvature
scale, $\lvert R^{(3)}\rvert^{-1/2}$, that comoves with the expansion.
 The combination of $00$ and $ii$ components characterizes
  the acceleration of the expansion.  To avoid an overly cumbersome notation, we omit overbars to denote these background values
 since perturbations are always designated as such below.  When referring to the resummed local
 background, we use the subscript $W$ for the locally windowed average below.

\subsection{Metric Perturbations}

Let us perturb the metric in an arbitrary gauge following Ref.~\cite{Bardeen:1980kt,Kodama:1985bj}. The most general metric perturbation we can write is of the form
\begin{eqnarray}
\delta g_{00} &=& -a^2 (2 {\cal A}), \nonumber \\
\delta g_{0i} &=& a^2 {\cal B}_i ,\nonumber \\
\delta g_{ij} &=& a^2 ( 2 {\cal E}_{ij}).
\end{eqnarray}

It is convenient to decompose these metric perturbations in two ways.  First we Hodge decompose the vector, ${\cal B}_i$, and tensor, ${\cal E}_{ij}$ perturbations:
\begin{align}
{\cal B}_i &= \nabla_i {\cal B}+{\cal B}^T_i,\\
{\cal E}_{ij} &=  {\cal E}_L \gamma_{ij}+\left(\nabla_{i}\nabla_{j} -\frac{1}{3}\gamma_{ij}\nabla^2\right){\cal E}_T+\nabla_{(i}^{\vphantom{T}}{\cal E}^T_{j)}+{\cal E}_{ij}^{TT}, \nonumber
\end{align}
where the vectors, ${\cal B}^T_i, {\cal E}^T_{j}$ are divergence free, and the tensor ${\cal E}_{ij}^{TT}$ is transverse and traceless. 
We are interested in absorbing the scalar metric perturbations into a local background and drop the divergence free and
transverse traceless pieces from consideration hereafter.

Second, it is helpful to make a harmonic decomposition of the scalar components of the metric.
The harmonics themselves are the complete and orthogonal set of eigenmodes of the spatial
Laplace operator
\begin{equation}
\nabla^2 Q({\bf x})= -k^2 Q({\bf x}).
\end{equation}
Orthogonality, or more fundamentally the homogeneity and isotropy of the background, means that we 
can consider each eigenmode independently for linear perturbations.
Because we consider one eigenmode at a time, we omit the ${\bf k}$ index for clarity.
These eigenmodes are plane waves when $K=0$, $Q = e^{i{\bf k}\cdot{\bf x}}$, and in this case our treatment reduces to the usual Fourier decomposition of a general spatial perturbation, examined one
mode at a time.  For details of the decomposition in the
$K\ne 0$ cases, see Ref.~\cite{Harrison:1967zza}.
To build the scalar component of ${\cal B}_i$ and ${\cal E}_{ij}$ it is useful to introduce
\begin{eqnarray}
Q_i ({\bf x}) &\equiv& - k^{-1} \nabla_i Q({\bf x}) ,\nonumber\\
Q_{ij}({\bf x}) &\equiv& \left(k^{-2}  \nabla_i\nabla_j  + \frac{1}{3}\gamma_{ij} \right)Q({\bf x}),
\end{eqnarray}
whose indices are raised and lowered by $\gamma_{ij}$, which also defines the
covariant spatial derivative  $\nabla_i$.  Our normalization conventions for $Q_i$ and $Q_{ij}$ are established to keep the harmonic space representation of metric perturbations dimensionless.  We then obtain for a single scalar eigenmode
\begin{eqnarray}
{\cal A}({\bf x},\eta)  &=&  A(\eta) Q({\bf x}), \nonumber \\
{\cal B}_i ({\bf x},\eta)  &=&
B(\eta) Q_i({\bf x}),\nonumber\\
{\cal E}_{ij}  ({\bf x},\eta)  &=&H_L(\eta) Q({\bf x}) + H_T(\eta) Q_{ij}({\bf x}),
\label{eq:metpertdecomp}
\end{eqnarray}
or equivalently $k {\cal B}({\bf x},\eta)= - B(\eta) Q({\bf x})$, ${\cal E}_L({\bf x},\eta)= H_L(\eta) Q({\bf x})$ and $k^2 {\cal E}_{T} ({\bf x},\eta)= 
 H_T(\eta)  Q({\bf x}).$
In order to relate these metric perturbations to the local expansion and curvature we next
consider their implications for the Einstein tensor. 
 
 \subsection{Einstein Tensor Perturbations}
 
With the decomposition~\eqref{eq:metpertdecomp}, the perturbation to the Einstein tensor can be written as
\begin{eqnarray}
\delta G^0_{{\hphantom 0}0} &=& \Pa Q ,\nonumber\\
\delta G^0_{{\hphantom 0}i} &=& \Pb Q_i , \nonumber\\
\delta G^i_{{\hphantom 0}j}   -\delta G^0_{{\hphantom 0}0} \frac{\delta^i_{{\hphantom i}j}}{3}&=& \Pc  Q \delta^i_{{\hphantom i}j}   + \Pd Q^{i}_{\hphantom{i}j},
\end{eqnarray}
where the individual components are given by
\begin{widetext}
\begin{eqnarray}
\label{eqn:Pnearlier}
\Pa  &=&  - \frac{2}{a^2}(k^2 - 3K)\left( {H_L}+ {{H_T} \over 3}  \right) + 6 H^2  A 
	 - 6 \frac{H}{a}  \left( \dot H_L  +   \frac{k B}{3} \right), \\
\Pb  &=& \frac{2k }{a^2} \Bigg[
 aH   {A} - \left( 1- \frac{3K}{k^2}\right)\left( {\dot H_L} 
+{{\dot H_T} \over 3 }   \right) 
 - {3K \over  k^2  } \left({\dot H_L} +\frac{k {B}}{3}   \right) \Bigg], \\
\Pc &=& \frac{2}{a^2} \Bigg[ \left( 2 \frac{\rd (a H)}{\rd\eta}  + aH  {\rd \over \rd\eta}
- {k^2 \over 3}\right) {  A}  
- \left( {\rd\over \rd\eta} + {\dot a \over a} \right) \left( {\dot H_L} + {k  {B} \over 3}  \right) \Bigg],  \\
\Pd &= & -\frac{1}{a^2} \Bigg[  k^2 \left(  {A} +  {H_L} + {{H_T} \over 3}   \right)
+ \left({\rd \over \rd\eta}+ 2 {\dot a \over a} \right)
	(k  {B} -   {\dot H_T}) \Bigg] .
	\label{eqn:Pn}
\end{eqnarray}
\end{widetext}
Each of these ${\cal G}$ components of the Einstein tensor contain up to second derivatives of the metric potentials.

\subsection{Gauge}
 
The general decomposition of the scalar components of the metric and Einstein tensor perturbations
in the previous sections applies to any gauge.   However, the values of the individual components  of course depend on the gauge and we shall see below there are special gauges for which the association of a local FRW background is the closest.  

Under a gauge transformation, or diffeomorphism, $x^\mu\mapsto x^\mu+\xi^\mu$, the scalar metric perturbations transform as
\begin{align}
\delta_\xi A &= - H (a T)' \nonumber\\
\delta_\xi B &= aH \left( L'+\frac{k}{aH} T\right),\nonumber\\
\delta_\xi H_L &= - \frac{k}{3}L- aH T,\nonumber\\
\delta_\xi H_T &= kL,
\label{eqn:metricgauge}
\end{align}
where we have split the diffeomorphism parameter as $\xi^\mu = (T, L Q^i)$. 
Here and throughout when  highlighting scalings, we employ dimensionless
e-fold derivatives
$' \equiv \rd/\rd\ln a =  (aH)^{-1}\rd/\rd\eta$.
 The variables $T$ and $L$
define the change in the time slicing and threading of the gauge, respectively.  Between
any two gauges that are fully fixed, $T$ and $L$ are uniquely defined.

Despite the fact that the metric potentials, as components of a tensor, manifestly transform under a diffeomorphism, and therefore take different numerical values in each gauge, we use the {\it same} symbols $\{ A, B, H_L, H_T \}$  or $\{ \Pa,\Pb,\Pc,\Pd \}$ to parameterize the metric and Einstein tensor in any gauge.  For example the lapse perturbation, $A$, in a  gauge that is completely fixed
 is a perfectly well defined geometric quantity when viewed as a spacetime object in its own right but  is not equal to the lapse perturbation in a different gauge.    
 
 To avoid confusion, when we specify relations that apply only to a specific gauge below, we will assign special variables to the components; for example we call the lapse perturbation in comoving gauge $\xi=A|_{\rm com}$ below. Since a gauge transformation between
 two fixed gauges is a one-to-one transformation that uniquely defines $T$ and $L$, 
a perturbation in one gauge may always be written as a unique combination of  the variables  in a different gauge, e.g.
\begin{equation}
\xi = A - H\left[ a T(A,B,H_L,H_T)\right]',
\label{eqn:lapseexample}
\end{equation}
where $T$ is specified in this case by Eq.~(\ref{eqn:Tcom}) below.
Combinations of variables of this type are often called ``gauge invariant" or ``Bardeen" variables in the
 literature.   
 Since the gauge-fixed and Bardeen variables, thought of as scalar functions in the spacetime, represent the same geometric objects---and  take on the same numerical values---we use the same special symbols for both, e.g.~$\xi$ in the example above.  We simply give
 $T$ and $L$ from which the Bardeen representation can be obtained with
 Eq.~(\ref{eqn:metricgauge}).
 
\subsection{Local Expansion and Curvature}
\label{sec:localcurvature}

We would now like to assign a geometric interpretation to the components of the 
Einstein tensor in the 3+1 decomposition.
In order to do this, we note that the covariant derivative of a 
unit  timelike vector
can in general be decomposed
with the help of the induced metric 
\begin{equation}
P_{\mu \nu } = g_{\mu \nu } + n_\mu n_\nu,
\end{equation}
 into the expansion
 \begin{equation}
 \theta \equiv \nabla_\mu n^\mu,
\end{equation}
  vorticity
 \begin{equation}
 \omega_{\mu\nu} \equiv P_\mu^{\hphantom{\mu}\alpha}P_\nu^{\hphantom{\mu}\beta}
(\nabla_\beta n_\alpha - \nabla_\alpha n_\beta) , 
\end{equation}
shear 
\begin{equation}
\sigma_{\mu\nu}  \equiv {1 \over 2} P_\mu^{\hphantom{\mu}\alpha}P_\nu^{\hphantom{\mu}\beta}
(\nabla_\beta n_\alpha + \nabla_\alpha n_\beta)  - {\theta \over 3}  P_{\mu\nu},
\end{equation}
and acceleration
\begin{equation}
a_\mu \equiv ( \nabla_{\alpha} n_\mu )n^\alpha
\end{equation}
 of the vector field such that
\begin{eqnarray}
\nabla_\nu n_\mu \equiv \omega_{\mu\nu} + \sigma_{\mu \nu} + {\theta \over 3}  P_{\mu\nu} - a_\mu n_\nu.
\end{eqnarray}

For the particular vector $n_\mu= (-a(1+AQ),{\bf 0})$ which is normal to constant time surfaces,
$\omega_{\mu\nu}=0$, $\sigma_{00}=\sigma_{0i}=0=a_0$
with (see e.g.~\cite{Kodama:1985bj})
\begin{eqnarray}
\sigma_{ij} &=& a^2 H \Sigma Q_{ij}, \nonumber\\
a_i &=& -k A Q_i, \nonumber\\
\theta  &=& 3 H (1 - A Q) + 3H \delta N' Q,
\label{eqn:expansion}
\end{eqnarray}
where the amplitude of the shear is
\begin{equation}
\Sigma  =
  H_T' - \frac{k}{aH}B,
\end{equation}
and we isolate particular terms in the expansion
\begin{eqnarray} 
\delta N' = H_L'+ \frac{k}{aH} \frac{B}{3} 
=  H_L' +\frac{H_T'}{3} - \frac{\Sigma}{3},
\label{eqn:deltaN}
\end{eqnarray}
 for reasons that we will now make clear  \cite{Wands:2000dp}.

We can now use these relations to interpret the components of the perturbed Einstein 
tensor, starting with $\Pa$.
In an FRW background the $\Pa$ component represents the metric side of the Friedmann equation. In order to generalize this to the perturbed case, we must define the local Hubble rate. Geometrically, the local volume expansion in an isotropic universe is  the fractional change per unit proper time, $\tau$,
of the cube of the local scale factor,
 $\rd\ln a_W^3/\rd\tau$, seen  on the worldline of an observer at fixed spatial coordinates. Therefore, we should define the local Hubble rate in a perturbed universe, $H_W$, using the expansion rate of the volume on spatial slices as
\begin{eqnarray}
H_W & \equiv & \frac{\rd\ln a_W}{\rd\tau}\equiv  \frac{1}{3} \theta  
 = H  +\delta H Q,
\end{eqnarray}
where we have defined
\begin{equation}
\delta H = (\delta N'- A)H  .
\end{equation}
Here and below the subscript ``$W$" denotes a local windowed average on scales much smaller than the
wavelength of the perturbation.
Note that this expression takes the form of the background Hubble parameter, plus a perturbation.
The term proportional to $A$ comes from the conversion from proper time of the local observer
through
 conformal time to cosmic time $\rd\tau = a(1+AQ) \rd \eta = (1+A Q)\rd t
$. The 
remaining piece can be attributed to the change in the  difference between the e-folds of the local and 
global expansion,
\begin{equation}
\ln a_W=\ln a+ {\delta N} Q,
\end{equation}
per unit cosmic time.  
Note that Eq.~\eqref{eqn:deltaN} defines $\ln a_W$ only up to a constant, 
which just amounts to choosing
a normalization epoch for the local scale factor. We choose this constant so that 
$a_W=a$ at some suitable initial epoch.

By defining $H_W$ in this more general way utilizing the volume expansion we also allow shear or
anisotropic perturbations, even though they do not exist in the FRW case.
Conversely, in these cases the separate universe ansatz is taken to mean that angle-averaged local observables are indistinguishable from those in the separate universe.    Furthermore, note that the observers that define the local
expansion in this section need not be geodesic observers, though in \S \ref{sec:curvature} we
use geodesic observers to define the separate universe condition.

Next we consider the perturbation to the scalar Ricci 3-curvature on this slicing
\begin{equation} 
\delta R^{(3)}  =\frac{4}{a^2} (k^2-3K)\left( H_L + \frac{1}{3} H_T \right) .
\end{equation}
In this sense $H_L+H_T/3$ is the potential for curvature fluctuations.

Combining the expressions for the perturbed Hubble parameter and the perturbed 3-curvature, we can write ${\cal G}_\rho$ as
\begin{equation}
\Pa =  - 6 H \delta H -\frac{1}{2} \delta R^{(3)} ,
\end{equation}
which combined with the background, Eq.~\eqref{eqn:Gbackground}, takes the locally resummed form
 $-3 H_W^2 - R_{W}^{(3)}/2$. Defining the local curvature by $K_W =a_W^2  R_{W}^{(3)}/6$ we can write the local Einstein tensor as
\begin{equation}
 \left.G^0_{{\hphantom 0}0} \right\rvert_W =
-3 \left( H_W^2 + \frac{K_W}{a_W^2} \right),
\label{eqn:perturbedG00sepun}
\end{equation}
where the local curvature is given explicitly by
\begin{eqnarray}
K_W 
&=&
 \left( \frac{a_W}{a}\right)^2 K+ \frac{a_W^2}{6}    \delta R^{(3)}    Q  \nonumber\\
& \equiv & K + \delta K Q ,
 \end{eqnarray}
which defines the local curvature fluctuation to linear order as
\begin{equation}
\delta K = \frac{2}{3}  (k^2- 3K) \left( H_L + \frac{ H_T}{3} \right)  + 2 {\delta N}K  .
\label{eqn:dK}
\end{equation}

Note the last  term is non-zero only if $K \ne 0$ and is associated with
the difference between a non-zero background curvature that comoves with the
global or local expansion.
If we take $k^2 \ll 3|K|$,
 both ${\delta N}$ and $H_L+H_T/3$, which we would consider a curvature potential in the opposite limit,    are themselves curvature perturbations, i.e.\
fractional changes to the background curvature, $K$.  In Ref.~\cite{Bertschinger:2006aw}, this association
was exploited to map a $k=0$ fractional change in the background curvature, $K$, onto the Newtonian gauge
potentials (see \S \ref{sec:MG}) in order to relate the dynamics of the perturbations for $k^2 \ll 3|K|$ to that of the background.
Since we seek to define the local expansion for wavelengths that are larger than the horizon but
smaller than the curvature scale, we do not utilize this approximation here.    Where no ambiguity should arise, we occasionally use the shorthand convention of the literature and call the curvature potential perturbation
 $H_L+H_T/3$ a curvature perturbation even for $k^2 \gg 3|K|$.

If the local expansion for a given slicing behaves as a separate FRW universe, 
 then the local curvature of this slicing is constant, $K_W' = {\delta K}' =0$.
 However
the geometric correspondence~\eqref{eqn:perturbedG00sepun} 
holds regardless of slicing or the validity of the separate universe ansatz. The form simply reflects geometric labels that we put on the components of
the Einstein tensor.  

Note that according to Eq.~\eqref{eqn:deltaN},
 the curvature potential changes  if
the e-folding rate, $\delta N'$, is not spatially uniform
or if there is shear in the expansion
 $\Sigma$.  If the shear is negligible then the change
in the curvature potential can be computed from the change in e-folds, a technique that is known as the
$\delta N$ formalism, often employed in the context of inflation~\cite{Wands:2000dp}.

Next let us interpret the  Einstein tensor component $\Pb$.  It has no analogue in a background FRW spacetime but
using Eq.~\eqref{eqn:dK} for the local curvature perturbation, we can rewrite it as
\begin{eqnarray}
\frac{\Pb}{H^2}  &=&  2\frac{  k  }{aH} \left(  A  - \frac{3}{2} \frac{{\delta K}'}{k^2} \right) .
\label{eqn:P1curv}
\end{eqnarray}
$\Pb$ takes $\delta K'$, the evolution of the local curvature on the slicing, and combines
it with the lapse perturbation $A$.
We shall see below that comoving gauge sets $\Pb=0$ to obtain a conservation law
for the comoving gauge curvature potential.

$\Pc$ is related to the perturbation to the local acceleration of the expansion. From the long wavelength
limit 
\begin{equation}
\lim_{k\rightarrow 0} \Pc
=  \frac{4}{a^2}  \frac{\rd (a H)}{\rd\eta} A + \frac{ 2 H}{a} \dot A - \frac{2}{a^2}
\left( \frac{\rd}{\rd\eta} +  \frac{\dot a}{a} \right) \dot {\delta N} ,
\label{eqn:P2limit}
\end{equation} 
 we can see that
\begin{equation}
 \left.G^i_{{\hphantom i}j} -\frac{1}{3} G^0_{{\hphantom 0}0} \delta^i_{{\hphantom i}j} \right\rvert_W\approx -  \frac{2}{a_W}  \frac{\rd^2 a_W }{\rd\tau^2}  \delta^i_{{\hphantom i}j}.
\end{equation}
Note that the limit in Eq.~\eqref{eqn:P2limit}
 involves dropping the $k^2 A$ term in $\Pc$.   If the lapse $A$ is set to zero
by a gauge choice, as in synchronous gauge, then there is no restriction on $k$. More generally, a sufficient condition for~\eqref{eqn:P2limit} to be true is
\begin{equation}
\left\lvert \frac{ \rd\ln A}{\rd\ln a} + 2 \frac{\rd \ln (aH)}{\rd\ln a} \right\rvert \gg   \frac{1}{3} \left( \frac{k}{aH} \right)^2 .
\end{equation}

The anisotropic $\Pd$ term, like the $\Pb$ term, does not exist for an FRW cosmology.  At the perturbative level, it takes the form
\begin{equation}
\frac{\Pd}{H^2}  = -\left(\frac{k}{aH} \right)^2 \left(A + H_L +\frac{H_T}{3}\right) + \Sigma' + \left( 3 + \frac{H'}{H} \right) \Sigma.
\end{equation}
  It is non-zero
 if there is a shear in the expansion, $\Sigma$, or---should the shear be set to zero as in Newtonian gauge below---if the lapse perturbation  is not equal and opposite to the
curvature potential.

As alluded to above, when fixing a gauge we will typically do it by specifying that one of these geometric quantities vanishes. 
It is therefore useful
to give their gauge transformation properties explicitly
\begin{align}
\delta_\xi \left(H_L+\frac{H_T}{3} \right) &= - aHT ,\nonumber\\
\delta_\xi (\delta N' ) &= -( aH T)' + \frac{1}{3}  \frac{k}{aH} kT ,\nonumber\\
\delta_\xi \left(\frac{\delta K'}{k^2}\right) &= -\frac{2}{3}( aHT)' +   \frac{2}{3}  \frac{KT}{aH}  , \nonumber\\
\delta_\xi \Sigma   &= -\frac{k}{aH} k T ,\nonumber\\
\delta_\xi \left( \frac{\delta H}{H}\right) &=  -  \left[ \frac{H'}{H} + \frac{1}{3} \left( \frac{k}{aH} \right)^2 \right] a H T,
\end{align}
and
\begin{align}
\delta_\xi \left( \frac{\Pa}{H^2} \right) &= 6\left[ \frac{H'}{H}-\frac{K}{a^2 H^2} \right] a H T,
\nonumber\\ 
\delta_\xi \left( \frac{\Pb}{H^2} \right) &= 2 \frac{k}{aH} \left( a H'  - \frac{K}{aH} \right) T ,\nonumber\\
\delta_\xi \left( \frac{\Pc}{H^2} \right) &= 2 \frac{(H^2+ H H')'}{H^2} a HT , \nonumber\\
\delta_\xi  \left( \frac{\Pd}{H^2} \right)  &=0.
\end{align}
Notice that these quantities, along with the lapse perturbation, $A$, whose gauge transformation is given in Eq.~\eqref{eqn:metricgauge},  depend only on the time slicing.  Therefore gauge conditions on these quantities define the time slicing through $T$ and generally no more than one of these quantities may be set to zero by a gauge choice.

\subsection{Effective Stress Tensor}
\label{sec:stress}

Many relationships in the literature for how the curvature potential evolves assume general relativity
and therefore are expressed in terms of the total stress energy of the matter.
In order to connect with this language but remove the assumption of general relativity, 
it is useful to also relabel  the Einstein tensor as an
effective stress tensor
\begin{equation}
 T_{\mu\nu} \equiv\frac{1}{8\pi G} G_{\mu\nu} ,
\end{equation}
so that at the background level $T^0_{{\hphantom 0}0} =-\rho$ and  $T^i_{{\hphantom 0}j}=p$ whereas
for the most general scalar perturbations
\begin{eqnarray}
\delta T^0_{{\hphantom 0}0} &=& -\delta\rho\, Q ,\nonumber\\
\delta  T^0_{{\hphantom 0}i} &=& (\rho+p) (v-B) Q_i , \nonumber\\
\delta  T^i_{{\hphantom 0}j} &=&\delta p\,  Q \delta^i_{{\hphantom i}j}   + p \,\pi \,Q^{i}_{\hphantom{i}j}.
\end{eqnarray}
Note that this reinterpretation does {\it not} require that the Einstein equations
hold, it is merely a convenient way of organizing the geometric object that is the Einstein tensor.

On the other hand, in general relativity these components correspond to familiar quantities
in the actual stress tensor of all matter species,
with $\delta \rho$ the  energy density fluctuation, $(\rho+p) (v-B)$
the momentum density, $p$ the  pressure fluctuation, 
and $p\, \pi$ the scalar anisotropic stress.

For modified gravity theories, where the Einstein equation
does not hold, we can view these effective stress energy components as
simply a convenient relabeling of the ${\cal G}$ linear combinations of the 
metric perturbations and their derivatives, leading to a dual set of interpretations
of their values as effective stress-energy components
\begin{eqnarray}
\Pa &\equiv& -8\pi G\, \delta \rho, \nonumber\\
\Pb &\equiv& 8\pi G (\rho+p) (v-B) ,\nonumber\\
\Pc &\equiv& 8\pi G (\delta p + \delta \rho/3), \nonumber\\ 
\Pd &\equiv & 8\pi G p\, \pi  .\label{eqn:metricstress}
\end{eqnarray}
Since $\nabla^\mu G_{\mu\nu}=0$ by virtue of the Bianchi identities, we can use the 
implied conservation equations $\nabla^\mu T_{\mu\nu}=0$
\begin{eqnarray}
\frac{[a^3 \, \delta \rho]'}{a^3} + 3\delta p 	&=&
        -(\rho+p)\left(\frac{kv}{aH}   + 3 H_L'\right),  \nonumber\\
\frac{aH}{k} \frac{[ a^4  (\rho + p)( {v}- {B})]'}{a^4}&=& 
{ \delta p }- {2 \over 3}\left( 1-{3K\over k^2}\right)p\,  {\pi} \nonumber\\
&&+ (\rho+ p)  {A} 
\label{eqn:conservation}
\end{eqnarray}
as dynamical relations between the metric components. These are ``conservation" equations for the effective stress components.
Substituting their metric definitions from Eqs.~\eqref{eqn:Pnearlier}--\eqref{eqn:Pn} and~\eqref{eqn:metricstress}, we see that these
relations are indeed identities and are automatically satisfied for the set of metric fluctuations
$\{ A,B,H_L,H_T \}$ in any gauge.

To move between gauges, it is useful to note that the effective stress components---or equivalently
the respective metric combinations, ${\cal G}$---transform as
\begin{align}
\delta_\xi (\delta \rho )&= - \dot\rho T,\nonumber\\
\delta_\xi (\delta p ) &= -\dot p T ,\nonumber\\
\delta_\xi v  &= \dot L ,\nonumber\\
\delta_\xi \pi  &= 0 .
\end{align}
Note that $\pi$ is gauge invariant in the full sense that the anisotropic stress takes the same value for all slicings and threadings.  These transformation properties also apply to any component of real stress energy in the universe as they follow from the transformation properties of a tensor.

Likewise if there is
a real component ``$a$" of stress energy that is separately covariantly conserved,
$\nabla_\mu T^{\mu\nu}_a=0$, then its energy and momentum densities obey the conservation laws~\eqref{eqn:conservation}.
In particular it will be useful in the next sections to consider a component of non-relativistic matter initially 
at rest with respect to the expansion.   In an arbitrary gauge,
its conservation equations become
\begin{eqnarray}
\delta_m' &=& - \frac{k v_m}{aH} - 3 H_L', \nonumber\\
\, [a(v_m-B)]' &=& \frac{k}{H} A,
\label{eqn:matterconservation}
\end{eqnarray}
where  $\delta_m=\delta\rho_m/\rho_m$, $p_m=0$ and $\pi=0$.
Note that this component need not be associated with real matter in the universe.  It could be a fictitious trace component
with $T^{\mu\nu}_m \rightarrow 0$ that has no impact on the expansion.  
Such a trace component simply defines a set of local observers.   
In this sense, all of our constructions here and in 
the next section are the same whether the metric is cast in Jordan or Einstein frame of a modified
gravity theory or whether the true matter is minimally coupled to the metric.   
In Einstein frame, where e.g.\ cold dark matter no longer falls on geodesics of the metric, the component $m$ is purely a device to establish a coordinate system of geodesic observers.

\section{Separate Universe and Curvature Conservation}
\label{sec:curvature}

In this section, we establish the conditions under which local angle-averaged observables in a
perturbed universe are
to good approximation  those of a separate FRW universe defined by the  local expansion and curvature---which we call the separate universe
ansatz. 
 These conditions take on various forms in various time slicings since in an exact FRW expansion
the preferred slicing is simultaneously synchronous, comoving, uniform density, uniform e-folding and zero
shear whereas in the perturbed universe, no single slicing can satisfy all of these properties simultaneously.

 The primary condition, as discussed in \S \ref{sec:synch}, is that the local curvature for synchronous
observers initially at rest with respect to the expansion is conserved.   In \S \ref{sec:com}, we show that this occurs
when synchronous and comoving gauge approximately coincide, satisfying the condition that the
comoving lapse perturbation is much smaller than the curvature potential.  
In \S \ref{sec:uniformdensity}, we discuss the relationship between these equivalent conditions and
the conservation of curvature on slicings of constant density or constant e-folding.

\subsection{Synchronous Gauge}
\label{sec:synch}

Local geodesic observers that are initially at rest with respect to the background expansion define a synchronous coordinate system.   
We use these observers to define the local FRW expansion.    Conservation of
the local curvature in this synchronous gauge is therefore the primary condition for the
validity of the separate universe ansatz.

In a synchronous gauge, the time reparameterization freedom is used to fix $A=0$ and the spatial gauge freedom is utilized to set $B=0$. 
This is the coordinate system defined by a set of geodesic observers that synchronize their clocks.  
There is residual gauge freedom in defining which observers establish the synchronous 
coordinates.     As discussed in \S \ref{sec:stress}, we can take  
this set of observers to be tracer particles of non-relativistic matter, $m$, initially at rest with 
respect the expansion and possessing spatially uniform density.  They subsequently fall on geodesics of the local metric and hence obey 
Eq.~\eqref{eqn:matterconservation}. Over a distance that is short compared with the wavelength of
the mode, synchronous coordinates coincide with comoving Fermi normal coordinates for an isotropic configuration (see e.g.~\cite{Dai:2015jaa}).

From an arbitrary gauge,
 we can reach this synchronous gauge with
\begin{eqnarray}
T = \frac{v_m - B}{k},\quad \dot L = -{v_m},
\end{eqnarray}
and Eq.~\eqref{eqn:matterconservation},
leaving a non-dynamical constant freedom in $L$  that is fixed by demanding that the spatial coordinates
are unperturbed initially.

Using these relations in the gauge transformation equations allows us to represent the synchronous gauge
perturbation variables in terms of the variables in an arbitrary gauge, i.e.\ the Bardeen representation.  
In particular, the matter velocity
in this gauge vanishes $v_m|_{\rm synch}=v_m + \dot L = 0$, but will not in an arbitrary gauge.

Let us give the curvature potential in this synchronous gauge a unique symbol
\begin{equation}
-\eta_T\equiv\left.H_L + \frac{H_T}{3}\right\rvert_{\rm synch},
\end{equation}
and define the remaining scalar  metric perturbation as $h_L\equiv\left.6 H_L\right\rvert_{\rm synch}$. 
From this point on in this section all quantities are evaluated in this synchronous gauge
unless otherwise specified.

  Using Eq.~\eqref{eqn:dK}, the
local curvature that the observers would define is given by $K_W =K+\delta K Q$ where
\begin{equation}
\delta K =  -\frac{2}{3}(k^2-3K)\eta_T + \frac{1}{3} K h_L .
\end{equation}
In order for the separate universe ansatz to hold exactly, we must have  ${\delta K}'=0$.   From Eq.~\eqref{eqn:P1curv} we 
can see that this condition is equivalent to requiring $\Pb=0$.    

From the association of the momentum term in the effective stress~\eqref{eqn:metricstress}, we see that this condition geometrically means that the effective velocity vanishes,
$v=0$. Since this is the gauge where $v_m=0$, the gauge invariant condition is
is that
the effective matter comoves with the synchronous matter: $v-v_m=0$.
  In other words, 
the validity of the separate universe approximation rests on whether synchronous gauge and comoving
gauge coincide.  Conversely, the local curvature evolves only if the effective matter moves away from the
geodesics that define the synchronous observers.

Moving away from this exact statement, we can define a condition under which the separate universe ansatz holds 
approximately, which only requires that the fractional change of $\delta K$  per e-fold  is small: 
$\lvert{\delta K'}\rvert\ll \lvert{\delta K}\rvert,$
so that
\begin{equation}
\left\lvert\frac {\delta K}{k^2} \right\rvert  \gg 
\left\lvert\frac{1}{3} \frac{a H}{k } \frac{\Pb}{H^2}\right\rvert = \left\lvert \frac{8\pi G (\rho+p)}{3H^2} \frac{a H}{k} v \right\rvert.
\label{eqn:syncgaugedeltaKcond1}
\end{equation}
When $K=0$, this reads
\begin{equation}
\left\lvert\eta_T\right\rvert \gg \left\lvert \frac{4 \pi G(\rho+p)}{H^2}\frac{a H}{k} v \right\rvert.
\label{eqn:synchronouscondition}
\end{equation}
Recall that the effective momentum conservation equation~(\ref{eqn:conservation}) provides the evolution equation for $v$
\begin{equation}
\frac{aH}{k} \frac{[ a^4  (\rho + p)v]'}{a^4}=
{ \delta p }- {2 \over 3}\left(1-3{K\over k^2}\right)p  {\pi},
\label{eqn:synchEuler}
\end{equation}
so we see that the conditions~\eqref{eqn:syncgaugedeltaKcond1} and~\eqref{eqn:synchronouscondition} restrict how much the 
effective stress gradients are allowed to generate momentum and move the effective matter off the
synchronous geodesics.   We will establish the relationship between this condition and the
equivalence of synchronous and comoving gauge next.

\subsection{Comoving Gauge}
\label{sec:com}

Comoving gauge is useful in that the separate universe condition can be phrased as simple 
algebraic relations rather than the pair of synchronous relations, Eq.~\eqref{eqn:synchronouscondition} and \eqref{eqn:synchEuler}.
 Since Eq.~\eqref{eqn:P1curv} for $\Pb$ involves $\delta K'$, taking comoving slicing where $\Pb=0$ gives an evolution equation for the curvature fluctuation.
 To get to comoving gauge from an arbitrary slicing, we apply the time shift
\begin{equation}
T =- \frac{\Pb}{2k}\left( HH' - \frac{K}{a^2} \right)^{-1}.
\label{eqn:Tcom}
\end{equation}
In this slicing, the effective momentum density vanishes, $T^0_{\hphantom{0}i}=0$, and 
hence  comoves with the coordinates, $v=B$  (see Eq.~\ref{eqn:metricstress}).  
Let us give  unique symbols for the  curvature potential and lapse
which are fully fixed by the slicing
\begin{equation}
{\cal R}\equiv \left.H_L+\frac{H_T}{3}\right\rvert_{\rm com}, \quad \xi \equiv \left.A\right\rvert_{\rm com}.
\end{equation}
From this point on in this section all quantities are evaluated on comoving slicing
unless otherwise specified.

  From Eq.~\eqref{eqn:P1curv} we obtain the curvature evolution equation 
\begin{equation}
{\delta K}'  = \frac{2}{3}  (k^2-3K) {\cal R}' + 2K {\delta N}' =  \frac{2}{3} k^2  \xi.
\end{equation}
The curvature on comoving slicing is conserved if the lapse is sufficiently small 
in comparison 
\begin{equation}
 | \xi | \ll |\delta K/k^2| .
 \label{eqn:curvaturecondition}
\end{equation}
We can alternately express this lapse condition in terms of the curvature potential
by eliminating 
 $\delta N'$ in favor of the shear
\begin{equation}
 {\cal R}' - \frac{K}{k^2} \Sigma = \xi.
\end{equation}
For $K=0$ this reduces to the simple condition that $|\xi| \ll |{\cal R}| $ and
we shall below  (see Eq.~\ref{eqn:lapseshear}) that assuming $|\xi| \ll |{\cal R}| $ generically 
implies $ |\Sigma| \ll |{\cal R}|$ above the horizon.   We therefore for convenience refer to the lapse
condition as 
\begin{equation}
\left\lvert\xi\right\rvert \ll \left\lvert{\cal R}\right\rvert \implies 
\left\lvert\frac{{\cal R}'}{{\cal R}} \right\rvert\ll 1\label{eqn:lapsecondition}
\end{equation}
for  $k^2 \gg \left\lvert K\right\rvert$ and $|K| \ll (aH)^2$, i.e.\ for all relevant scales
to be below the background curvature scale.
More generally Eq.~(\ref{eqn:curvaturecondition}) is the  direct and precise statement
and does not require any condition on $K$.

The utility of working in comoving slicing is that we can simply state the condition for curvature conservation as algebraic relationships between the metric variables.
To further establish the connection with the separate universe condition,
we can specify the threading to fully fix the gauge, even though it does not enter into the lapse condition.
A convenient choice is  ``comoving threading,"  so that $G^i_{\hphantom{i}0}=0$.
In this case $v|_{\rm com}=0$ and we can reach this gauge from a gauge with finite $v$ by a diffeomorphism with $\dot L = -{v}$.  
This does not uniquely fix the threading since it allows an arbitrary time-independent diffeomorphism
 $L= $\,const. which shifts the coordinates by $\delta x^i \propto Q^i(x)$,
 but  we again fix this ambiguity by taking the coordinates to be unperturbed initially.

When the lapse condition Eq.~\eqref{eqn:lapsecondition} is satisfied, this fully-fixed comoving gauge approximately coincides with
synchronous gauge.
More explicitly, defining synchronous observers as in \S \ref{sec:synch}, following the same geodesics as freely falling matter,
the synchronous gauge curvature is given in comoving variables by
\begin{equation}
-\eta_T = {\cal R} - \frac{a H}{k} v_m,
\end{equation}
and matter conservation, Eq.~\eqref{eqn:matterconservation}, gives
\begin{equation}
v_m
= \frac{1}{a}\int \frac{\rd a}{a} \frac{k}{H} \xi ,
\end{equation}
where the condition that the matter is initially at rest fixes the integration constant.
Combining these equations, we see that the curvatures in the two gauges are the same, $-\eta_T \approx {\cal R}$,
when the comoving lapse condition $|\xi |\ll |{\cal R}|$ is satisfied.

We emphasize that in general comoving slicing and synchronous slicing will {\it not} coincide, since $\Pb=0$ and $A=0$ both define the time slicing. 
However it is precisely the case where they approximately coincide due to the comoving lapse condition that the separate universe
ansatz holds.

The condition~\eqref{eqn:lapsecondition} is nice because it is a purely geometric test for when the comoving curvature will be conserved (and hence, when the separate universe ansatz holds). However, it is sometimes helpful to think in terms of the effective stresses on the metric, so we can also express the  lapse condition  in terms of algebraic constraints on the effective stress tensor.   In particular the momentum conservation equation \eqref{eqn:conservation} gives an algebraic
relation for the comoving gauge lapse in terms of the stresses \cite{Bardeen:1980kt}
\begin{equation}
({\rho+p})\xi = -{\delta p} + \frac{2}{3}\left(1- \frac{3K}{k^2}\right) p\,\pi.
\label{eqn:lapsestress}
\end{equation}
Thus the small lapse condition can be re-expressed as a comparison between the curvature
perturbation and the stress perturbations in comoving slicing
\begin{equation}
{\cal R}'  = - \frac{\delta p}{\rho+p} + \frac{2}{3} \frac{p\,\pi}{\rho+p} .
\label{eqn:Rprimestress}
\end{equation}
for $k^2 \gg 3|K|$.
In particular, on large scales $k/aH \ll 1$, the comoving curvature is still conserved even if
there is an effective anisotropic stress $p\,\pi$ that is of order the isotropic
stress fluctuation $\delta p$ as long as both are suppressed by powers of $k/aH$ compared
with the curvature (cf.~\cite{Dai:2015jaa}).  This also includes any contribution from
non adiabatic pressure, $\delta p - (p'/\rho')\delta\rho$.
Furthermore since $p \, \pi$ is the source of shear in the expansion $\Sigma$ through
Eq.~\eqref{eqn:metricstress}, the comoving lapse condition \eqref{eqn:lapsecondition} also generally implies
\begin{equation}
\lvert\Sigma \rvert \ll \lvert {\cal R} \rvert ~~~~~{\rm for} ~~~~~k \ll aH.
\label{eqn:lapseshear}
\end{equation}
We shall see next that although the lapse condition \eqref{eqn:lapsecondition} 
generically implies the shear condition \eqref{eqn:lapseshear} the converse is not true and leads to another 
perspective on the conservation of curvature above the horizon.

\subsection{Uniform Density Gauge}
\label{sec:uniformdensity}

A related, but fundamentally different, criterion for the conservation of curvature can be obtained from 
the uniform density gauge 
of a real density component ``$a$"  whose stress energy is separately conserved: $\nabla_\mu T^{\mu\nu}_a=0$.   If this component has a barotropic equation of state $p_a(\rho_a)$, then choosing a gauge where  $\delta \rho_a=0$ sets
 $\delta p_a=0$ as well.  The continuity equation \eqref{eqn:conservation} then becomes an 
evolution equation for the curvature 
\cite{Wands:2000dp}.

For simplicity and definiteness let us assume that the component is the nonrelativistic matter of
 Eq.~\eqref{eqn:matterconservation}, $a=m$.   Again this need not be a component of real matter that impacts the expansion.   It merely establishes the coordinate system.

The spatially uniform 
density condition defines the slicing of the gauge and the gauge transformation from any other gauge is given by
\begin{equation}
T= \frac{\delta \rho_m}{\dot \rho_m} = -\frac{1}{aH} \frac{\delta_m}{3}\,.
\end{equation} 
In order to fix the threading of the gauge, we set $H_T=0$ and call the curvature potential in this gauge $\zeta_m$, the 
lapse perturbation $A_m$, and the shift perturbation $B_m$.  The
curvature in this gauge is related to the curvature and matter density perturbations 
in an arbitrary gauge as
 \begin{equation}
{\zeta}_m = H_L + \frac{H_T}{3}  +\frac{\delta_m}{3}.
\label{eqn:zetam}
 \end{equation}
The curvature potentials
 only agree for conditions and gauges where the density fluctuation is
already small compared with the curvature potential.  Then the shift to constant density slicing
does not significantly change the curvature. 
From this point on in this section all quantities are evaluated in uniform density 
gauge unless otherwise specified.

The benefit of this gauge is that the matter continuity equation \eqref{eqn:matterconservation} provides the evolution equation for
the curvature
\begin{equation}
\zeta_m' = -\frac{k}{aH} 
\frac{v_m}{3} ,
\end{equation}
so that the condition for conservation becomes 
\begin{equation}
\left\lvert\zeta_m\right\rvert \gg \left\lvert \frac{k}{aH} \frac{v_m}{3}  \right\rvert.
\label{eqn:constantdensitycondition}
\end{equation}
Though this condition \eqref{eqn:constantdensitycondition} can be stated independently of
the Einstein equation, its domain of validity in $k$ cannot without knowing the dynamics that sets the relative amplitudes of $\zeta_m$ and $v_m$.   In this sense it is no different than
the analogous synchronous gauge condition
\eqref{eqn:synchronouscondition} or the comoving lapse condition  \eqref{eqn:lapsecondition}.

To probe the differences with the other conditions, we can again examine
the momentum conservation equation \eqref{eqn:matterconservation}
\begin{equation}
v_m   = \frac{1}{a} \int \frac{\rd a}{a} \frac{k}{H} A_m + B_m,
\end{equation}
so that
\begin{equation}
\zeta_m' = -\frac{k}{a^2 H}  \int \frac{\rd a}{a} \frac{k}{H} A_m -\frac{k}{aH} B_m.
\end{equation}
For $\lvert\zeta_m'/\zeta_m\rvert \ll 1$,  both the lapse, $A_m$, and the shift, $B_m$, must be
in some sense small.   On the other hand,
 their amplitudes can be as large as the curvature itself, $\zeta_m$, 
and still allow $\lvert\zeta_m'/\zeta_m\rvert \ll 1$ as $k/aH \rightarrow 0$.   

This condition on the metric is apparently weaker than the comoving lapse or 
synchronous conditions.   
For example in a multi-fluid system with isocurvature modes, the curvature perturbations,
$\zeta_a$, in the constant density gauges of each component, $a$, are conserved outside the horizon even when the comoving curvature ${\cal R}$ is not \cite{Wands:2000dp}.

More explicitly, in our context the 
  conservation of $\zeta_m$ does not necessarily imply conservation of either ${\cal R}$ or
 $\eta_T$.
To see this, we can take the derivative of Eq.~(\ref{eqn:zetam}) and use 
the matter continuity equation (\ref{eqn:matterconservation}) in an arbitrary gauge
\begin{eqnarray}
\zeta_m' & =& -\frac{k}{aH}\frac{v_m}{3}+ \frac{H_T'}{3} \nonumber\\
&=& -\frac{k}{aH}\frac{v_m-B}{3} + \frac{\Sigma}{3}.
\end{eqnarray} 
This equation does not relate the evolution of the curvature perturbation in any other gauge to the evolution of $\zeta_m'$ precisely because $H_L'+H_T'/3$ drops out of the right hand side. 
Thus, we see that conservation of the curvature
in constant density gauge as $k/aH \rightarrow 0$ relies on the condition that the
shear is much less than the curvature in amplitude $\lvert\Sigma\rvert \ll \lvert\zeta_m\rvert$.

In the previous section we showed that the lapse condition generally implied the
shear condition above the horizon.   Since the converse is not necessarily true, the shear condition alone does not establish the separate universe condition for geodesic observers.
For example when  $\lvert\Sigma\rvert \ll \lvert\zeta_m\rvert$ but $|A_m| \sim | \zeta_m|$, the local FRW coordinate system constructed
from synchronous observers would still violate the separate universe criteria.  
Using the gauge transformation to synchronous gauge from constant density
gauge and the continuity equation~\eqref{eqn:matterconservation} we can obtain the evolution equation for the synchronous gauge curvature:
\begin{eqnarray}
-\eta_T' &=& \zeta_m' -\left[ \frac{a H}{k}  (v_m -B_m)\right]'\nonumber\\
	&=& \zeta_m' - A_m - H' \int \frac{\rd a}{a} \frac{A_m}{H},  
\end{eqnarray}
so that $\eta_T$ can evolve significantly even if $\zeta_m$ does not as $k/aH \rightarrow 0$ if
$\lvert A_m\rvert \sim \lvert\eta_T\rvert $.
For the two conditions to coincide, it must be the case that $\lvert A_m\rvert \ll \lvert\zeta_m\rvert$ which is again a condition on the lapse rather than on the shear.

In fact, we can interpret and generalize the relationship between curvature and shear by recalling that Eq.~\eqref{eqn:deltaN}
also provides an evolution equation for the curvature potential in terms of the change
in e-folds $\delta N'$ and the shear in an arbitrary gauge
\begin{equation}
H_L' + \frac{H_T'}{3} = \delta N' + \frac{\Sigma}{3}.
\end{equation}
Thus in a uniform e-folding gauge, where $\delta N'=0$, the curvature only evolves if there is shear in the
expansion.    Uniform density gauge for any matter component with a barotropic equation 
of state is a constant e-fold gauge as $k/aH \rightarrow 0$, if the density only evolves
because of the change in the spatial volume rather than the momentum density of the matter. 
Observers on fixed spatial coordinates  will then see a local expansion
that conserves the local curvature.  
However, since the shear condition does not
guarantee they are geodesic observers, we
do not consider uniform e-folding gauges further.

\section{Separate Universe Scale}
\label{sec:scales}

The comoving lapse condition, or the equivalence of synchronous and comoving gauge, provides
general conditions for the conservation of curvature and the validity of the separate universe
ansatz but does not directly provide a physical scale above which they are satisfied.
   To get further insight on
this question, there are two general approaches one can take.   The first is to parameterize relationships between the effective stress-energy 
and metric which we discuss in \S \ref{sec:Jeans}.  This has the benefit of generality, as it
applies to cases where the matter is non-minimally coupled and reduces to well-known results in
general relativity when the effective stress-energy is the actual matter stress-energy.   The second, which we discuss in \S \ref{sec:MG}, is to parameterize relationships between the actual stress-energy of matter and the metric,
i.e.~the modifications to the Einstein equations applicable to wide class of modified gravity models
\cite{Caldwell:2007cw,Hu:2007pj,Hu:2008zd,Hojjati:2011ix,Zhao:2008bn,Gleyzes:2013ooa,Gleyzes:2014rba,Baker:2015bva}.

\subsection{Comoving Jeans Scale}
\label{sec:Jeans}

The separate universe condition \eqref{eqn:lapsecondition} expressed in comoving gauge is that the lapse perturbation
must be much smaller than the curvature potential.    The lapse itself is directly related to the effective
stresses through Eq.~\eqref{eqn:lapsestress}.   We therefore need to make a connection between the
curvature and the effective stress-energy to close the system and determine the scale or domain of validity of the separate universe ansatz.  In this section we express all perturbations
 in terms of the comoving gauge quantities.

Using the $\Pa$ effective stress energy equation \eqref{eqn:metricstress}, we have for
$k^2\gg 3\lvert K\rvert$ and $a^2H^2 \gg \lvert K\rvert$,
\begin{eqnarray}
{\cal R} & \approx & 3\left( \frac{aH}{k} \right)^2 \left( {\cal R}' - \delta N' + \frac{1}{2}\frac{\delta \rho}{\rho}  \right) \nonumber\\
&=&  \left( \frac{aH}{k} \right)^2 \left(  \Sigma +\frac{3}{2} \frac{\delta \rho}{\rho} \right).
\label{eqn:RGrho}
\end{eqnarray}
We can eliminate $\delta \rho$ by defining the sound speed on comoving slicing as
\begin{equation}
c_s^2 \equiv \frac{\delta p}{\delta \rho},
\end{equation}
and relate it to ${\cal R}'\approx \xi$ by defining an analogous ``Jeans" speed that includes anisotropic stress
\begin{equation}
c_J^2 \equiv \frac{\delta p- 2p\pi/3}{\delta \rho} = -\frac{(\rho+p) {\cal R'}}{\delta\rho}.
\end{equation}
Finally, we can eliminate $\Sigma$ by taking the derivative of Eq.~\eqref{eqn:RGrho} and using
 the 
 $\Pc$ effective stress energy equation \eqref{eqn:metricstress} in the form
\begin{eqnarray}
\frac{\Sigma'}{3} &\approx& \frac{4\pi G}{H^2}\left(\delta p + \frac{\delta \rho}{3}\right)-  \left(2+\frac{H'}{H} \right) \frac{\Sigma}{3}  \nonumber\\
&&{-} \left[ \frac{H'}{H}-\frac{1}{3} \left(\frac{k}{aH}\right)^2\right] {\cal R}'
\label{eqn:RGp}
\end{eqnarray}
to obtain the equation of motion for ${\cal R}$
\begin{equation}
\frac{1}{a^3 } \left( a^3 H'  \frac{{\cal R}'}{c_J^2} \right)' \approx 3H' {\cal R}'
\left(1- \frac{c_s^2}{c_J^2} \right) - \left( \frac{k}{aH}\right)^2 {H'}{\cal R},
\label{eqn:Reom}
\end{equation}
where the only approximation is a negligible background curvature:
$k^2\gg 3\lvert K\rvert$ and $a^2H^2 \gg \lvert K\rvert$.   In particular since $c_s^2$ and $c_J^2$ are relations
for the effective stress, they supply closure relations in terms of the metric---not the true matter---and hence
do not assume the validity of the Einstein equations.  From this equation, we can read off the
regimes where $\lvert{\cal R}'/{\cal R}\rvert \ll 1$ is a solution.   

For the case of negligible effective anisotropic stress, $c_J^2=c_s^2$ and 
we can formally integrate Eq.~(\ref{eqn:Reom})  to find
\begin{equation}
{\cal R}' =- \frac{c_s^2}{a^3 H'}\left[  \int \frac{\rd a}{a} a^3 \left(\frac{k}{aH}\right)^2 H' {\cal R} +{\rm const.}\right].
\label{eqn:Rsoln}
\end{equation}
There are two generic cases of interest.  If the integrand is growing with $a$ then the integral term will be
dominated by the last few e-folds and its contribution to ${\cal R}'$ will scale as $\sim(c_s k/aH)^2 {\cal R}$.  
If the integrand receives its contribution mainly from earlier epochs then the integral will be constant
or at most logarithmically growing with $a$ and play a role similar to the constant of integration term.

In both cases, if $\lvert c_s^2/a^3 H'\rvert$ decreases with $a$ as $a^{-p}$ with $p>0$ then
${\cal R}$ will be nearly constant over the e-fold time scale
if $\lvert c_s k/aH\rvert \ll 1$.   This is the normal case where the comoving curvature is conserved outside the
sound horizon.  If $p\le 0$ then ${\cal R}'$ can grow with $a$, allowing ${\cal R}$ to change significantly
outside the sound horizon.
   This phenomenon occurs
in  inflation when the background is rapidly approaching de Sitter on a nearly
flat potential \cite{Kinney:2005vj} and violates the separate universe condition and hence the
 consistency relation between the power spectrum and bispectrum
\cite{Namjoo:2012aa}.  Other cases of a growing $1/a^3 H'$ were given in Ref.~\cite{Martin:2012pe} 
where there is  excess kinetic energy in the field beyond that expected from the local slope of the potential
on the attractor.

In the case where anisotropic stress dominates, $\lvert c_J^2 \rvert \gg \lvert c_s^2\rvert$.  In this situation the first term
on the right hand side of Eq.~\eqref{eqn:Rsoln} contributes.    If $|c_J^2| \ll1$ then the arguments 
above for conservation of curvature above the 
sound horizon apply with the generalization to the Jeans horizon. That is, ${\cal R}$ is conserved for $\lvert c_J k/aH \rvert\ll 1$.   If $\lvert c_J^2\rvert \gg 1$
then the additional term can be larger and limit the scale where ${\cal R}$ is conserved to the horizon, $k/aH \ll 1$.   In this
case the anisotropic stress dominates and makes the $\delta N'$ and $\delta \rho/\rho$ terms in  Eq.~\eqref{eqn:RGrho} negligible in comparison to ${\cal R}'$.

It is also important to note that although $c_s^2$ and $c_J^2$ play the role of closure relations, similar
to equations of state for the matter in general relativity with a single matter fluid, in
general they need not have any correspondence to the true matter equations of state, nor are
they specified by matter content and background alone.   
For example even in Einstein gravity
$c_s^2$ is defined as the ratio of the total pressure perturbation to the total energy
density perturbation on comoving slicing, which is defined by the total matter.   
In a multi-fluid case, this does not correspond to
the sound speeds of the individual components. 
In the presence of isocurvature modes, $c_s$ defined in this way can be very large, 
 $c_s^2  \gg 1$, 
because the energy densities of the multiple components can cancel, and
 ${\cal R}$ can evolve by a significant amount
arbitrarily far outside the horizon.  

Another interesting case is that of the cuscuton model, where the kinetic
term of the k-essence field in the Lagrangian carries no energy density fluctuations \cite{Afshordi:2007yx}.  Hence in the time slicing where
the field is spatially uniform, or equivalently comoving slicing with respect to the field, the finite pressure fluctuation and zero energy density fluctuation implies
an infinite sound speed. However, in the presence of other matter components, the sound speed of the total matter remains finite if there is
an additional normal matter component and one can again define a comoving sound horizon above which ${\cal R}$
is constant.   The comoving sound horizon therefore grows beyond the horizon as the universe enters late-time acceleration due to field domination.

In both the flat 
roll and cuscuton  examples,  we can trace the origin of the violation of the separate
universe condition to the more
direct relation \eqref{eqn:Rprimestress} and the simple and general criteria that the lapse perturbation
should be smaller than the curvature potential Eq.~\eqref{eqn:lapsecondition}.  Since 
 \begin{equation}
8\pi G (\rho+p) = -2 HH' + 2\frac{K}{a^2}
\end{equation}
the impact of stress fluctuations on the evolution of the comoving curvature is enhanced
if the expansion rapidly approaches de Sitter at negligible background curvature. 

In models where the Einstein equations no longer hold between the true matter and the metric these
definitions still hold but $c_s^2$ and $c_J^2$ depend on the modification to gravity and hence need not bear
any relationship to the equations of state of the matter components alone.

\subsection{Parameterized Gravity}
\label{sec:MG}

A second approach to developing a sense of scale for the separate universe beyond general relativity
 is to parameterize modifications through relations between the true matter stress energy and the metric variables.  This approach usually assumes that the true matter is minimally coupled to the metric and so we restrict ourselves to minimal coupling in this section alone. 
 
Parameterizations of this type are most commonly done in 
conformal Newtonian gauge, so we begin by reviewing
its properties.   
  In conformal Newtonian gauge we take a shear-free slicing, $\Sigma=0$, and isotropic threading: specifically
$A\equiv\Psi$, $B=0$, $H_L\equiv\Phi$ and $H_T=0$. The diffeomorphism parameters to get to conformal Newtonian gauge from an arbitrary gauge are
\begin{equation}
T= \frac{ aH}{k^2} \Sigma, ~~~~\quad L = -\frac{H_T}{k}.
\end{equation}
From this point on in this section all quantities are evaluated in this conformal Newtonian gauge
unless otherwise specified.

In this gauge the anisotropic component of the Einstein tensor is
\begin{equation}
\Pd = -\left(\frac{k}{aH}\right)^2 (\Phi+\Psi).
\end{equation}
Consequently, the first parameterized modification to the Einstein equation, motivated by scalar tensor theories which change the ratio $\gamma \equiv -\Phi/\Psi \ne 1$, even in the absence of matter anisotropic stress, is
to allow this relationship to be general $\gamma(a,k)$.  Note that in the effective stress language, this 
means that there is an effective anisotropic stress that is parameterized in terms of the metric as
\begin{equation}
8\pi G p\,\pi =  -\left(\frac{k}{aH}\right)^2 (1-\gamma)\Psi,
\end{equation}
 which is often
called the gravitational slip \cite{Caldwell:2007cw}.

This generalization does not in and of itself restrict the separate universe criteria but
rather restricts the evolution of $\Phi$ and $\Psi$, given that of $\gamma$.   To see this, we can write the
comoving curvature in Newtonian gauge variables, where it takes the form
\begin{equation}
{\cal R} = \Phi + H^2   \left( \Psi -\Phi' \right)\left( HH'- \frac{K}{a^2} \right)^{-1} \,.
\end{equation}
In a spatially flat $(K=0)$ background, which we will assume for the rest of this section,
 this becomes 
\begin{equation}
{\cal R} = \Phi +\frac{H}{H'}({\Psi - \Phi'}).
\end{equation}
If we now take the curvature to be conserved, ${\cal R}'=0$, we obtain a consistency relation between the metric potentials \cite{Hu:1998tj}
\begin{equation}
\Phi'' - \Psi' - \frac{H''}{H'}\Phi' - \left( \frac{H'}{H} - \frac{H''}{H'} \right) \Psi = 0.
\label{eqn:phipsi}
\end{equation}
Note that the compatibility of gravitational slip with the separate universe criteria can be seen directly
in comoving gauge.  Recall that an effective anisotropic stress $\pi/{\cal R} \sim (k/aH)^2$ did not violate
the separate universe criteria.  For example in general relativity, in the radiation dominated epoch 
$\gamma \ne 1$, due to neutrino anisotropic stress and yet we can take inflationary curvature predictions at horizon 
exit as fixed outside the horizon.  Conversely it is not correct to say that the curvature
conservation allows gravitational slip but strictly forbids anisotropic stress (cf.~\cite{Bertschinger:2006aw,Dai:2015jaa}).

The second parameterization involves the relationship between the true matter variables and the metric. 
Consider again the case where the true matter is non-relativistic, initially at rest with respect to the expansion, and falls on geodesics of the metric.  The matter then defines a synchronous coordinate system
and its curvature can be written in Newtonian variables as
\begin{equation}
-\eta_T = \Phi - \frac{a H}{k} v_m.
\label{eqn:etaTphi}
\end{equation}
If the  synchronous curvature is conserved, $\eta_T'=0$, then Eq.~\eqref{eqn:phipsi} also holds 
by differentiating  Eq.~\eqref{eqn:etaTphi} and employing momentum
conservation of the matter
\begin{equation}
(a v_m)' = \frac{k}{H}\Psi.
\label{eqn:Newtonianvm}
\end{equation}
Of course, these are equivalent because conservation of either $\eta_T$ or ${\cal R}$ implies
that synchronous and comoving gauge coincide.  In fact 
in the modified gravity literature \cite{Hu:2007pj,Hojjati:2011ix,Baker:2015bva}, the curvature on synchronous or matter comoving slicing is often denoted ${\cal R}$ or $\zeta$ when radiation is negligible.   With radiation, it is common to employ these variables  as the curvature on comoving slicing of the total matter \cite{Hu:2008zd,Zhao:2008bn}.  The way the we have defined ${\cal R}$ here only coincides with these alternate definitions when
the separate universe condition applies.  

Note further that in terms of the effective velocity, $v$, the comoving gauge potentials
can also be written in Newtonian variables as
\begin{eqnarray}
{\cal R} & = &\Phi - \frac{a H}{k}v, \nonumber\\
\xi &=& \Psi - \frac{H}{k} (a v)'.
\end{eqnarray}
When the separate universe condition holds,  $v\approx v_m$ and so, using Eq.~\eqref{eqn:Newtonianvm}, $\xi \approx 0$.   Any relationship between the Newtonian potential $\Psi$ and the matter can be made compatible with
the comoving lapse condition \eqref{eqn:lapsecondition} as long as this holds.

Therefore we would like to see under what conditions $v\approx v_m$.   By definition
\begin{equation}
 v \equiv \frac{k}{aH'}(\Phi' -\Psi);
\label{eqn:SUvm}
\end{equation}
to relate this quantity to $v_m$, let us define the synchronous---or matter comoving gauge---density perturbation 
$\Delta_m =\left.\delta_m\right\rvert_{\rm synch}$ using the Newtonian
gauge density variables
\begin{equation}
\Delta_m \equiv \delta_m + 3\frac{ aH}{k} v_m.
\end{equation}
Combining these equations with the matter conservation equations, we obtain
\begin{eqnarray}
v_m &=& \frac{k}{a H'} \left( 1- \frac{k^2}{3 a^2 H H'} \right)^{-1} \left( \Phi'-\Psi + \frac{\Delta_m'}{3} \right).
\end{eqnarray}
In order for the separate universe condition to hold, we must have $v_m\approx v$. Sufficient conditions for this to be the case are
\begin{equation}
k^2 \ll a^2 H H' ,\qquad \left\lvert \Phi' - \Psi \right\rvert \gg \frac{1}{3}  \Delta'_m,
\label{eq:confgaugevmisv}
\end{equation}
if the matter is separately conserved.    The latter condition is typically implemented 
phenomenologically by relating $\Delta_m$ and $\Psi$ through a ``modified Poisson equation"
\cite{Caldwell:2007cw,Hojjati:2011ix,Zhao:2008bn,Baker:2015bva}
\begin{equation}
k^2 \Psi = -4\pi G\mu  a^2 \rho_m  \Delta_m,
\end{equation}
where $\mu(a,k)$ parametrizes the general relation.  In this case the 
second condition~\eqref{eq:confgaugevmisv} is generally satisfied if 
\begin{equation}
\left( \frac{k}{aH} \right)^2 \left( \frac{3 H^2}{8\pi G \rho_m\mu } \right){\rm max}\left( \Big| 1 - \frac{\mu'}{\mu} \Big|,\frac{\Psi'}{\Psi}\right) \ll 1
\end{equation}
unless $\Phi' \approx \Psi$.
In this sense, conservation of curvature as $k\rightarrow 0$ requires only
matter conservation and very mild assumptions about how matter density sources
metric fluctuations.  

Although the $\mu,\gamma$ parameterization is very general, just like the sound speeds
of the effective stress in the previous section, the functions are in general determined not just 
by the background but by the solutions of the perturbation equations themselves.
For example, in the case where dark energy is an additional scalar field with a sound horizon,
$\mu(a,k)$ encodes the effect of its stress energy on the metric.    

There is also a hybrid approach that merges the effective stress and parametrized
gravity approaches.   In this case the effective stress
is formally separated into a ``dark energy" component ``$e$" and the normal matter components ``$a$"
\begin{eqnarray}
T^{\mu\nu}_{e} \equiv T^{\mu\nu}-\sum_a T^{\mu\nu}_a = \frac{G^{\mu\nu}}{8\pi G}-\sum_a T^{\mu\nu}_a.
\end{eqnarray}
Though this is fully general for matter whose joint stress-energy tensor is covariantly conserved, the  conservation equations $\nabla_\mu T^{\mu\nu}_{e}=0$ implied by the Bianchi identities require closure relations to complete.
One approach to these closure relations is to construct them to return $\mu$ and 
$\gamma$ for modified gravity  \cite{Hu:2007pj,Hu:2008zd} or mimic those of an actual physical component of dark energy
\cite{Fang:2008sn}.

Another approach is to 
parameterize the modifications to gravity at the level of the lagrangian
of an effective field theory or ADM formalism for the scalar-tensor perturbations  \cite{Creminelli:2008wc,Park:2010cw,Gubitosi:2012hu,Bloomfield:2012ff,Kase:2014cwa,Gleyzes:2014rba}.  In this language, conservation
of the  curvature in unitary or uniform scalar field gauge as $k \rightarrow 0$ is enforced by the structure of the lagrangian---or equivalently,
the implied closure relations for the  stress-energy of the effective dark energy component.  Unitary gauge and comoving gauge differ in models with kinetic braiding \cite{Sawicki:2012re} but the gauges do coincide for adiabatic fluctuations
as $k\rightarrow 0$. We defer a treatment of the scale associated with these
asymptotic behaviors to a future work \cite{MotHu16}.

In all cases and parameterizations, the primary consideration for the validity of the separate universe ansatz is that
lapse in comoving slicing is smaller than the curvature or equivalently the
effective matter comoves with synchronous observers.

\section{Discussion}
\label{sec:discussion}

We have established purely geometric or metric based criteria for the validity of the separate
universe ansatz.   Based on identifying the local volume expansion and curvature in a long-wavelength
perturbation with the global quantities for the background of a homogeneous
separate universe,  we have shown that the criterion is that the lapse is much smaller in 
amplitude than the curvature  potential on comoving slicing where the Einstein tensor $G^0_{\hphantom{0}i}=0$.  In this case, the Einstein tensor when interpreted as an
effective stress-energy tensor ``comoves" with the freely falling synchronous
observers which establish the local expansion so that the local curvature is conserved. 

In general relativity, this condition reduces to the familiar notion that the total matter
comoves with synchronous observers if only gravitational forces act.  
The lapse condition therefore is equivalent to the statement
that comoving gauge  and synchronous gauge coincide.  This condition allows anisotropic
local expansion of the volume or equivalently anisotropic effective stress that is as large
as their isotropic counterparts as long as they are both smaller than the curvature potential.  
In this case, the local universe behaves like a separate universe for angle-averaged
observables. 

The comoving lapse condition provides a general prescription for when the evolution of
small-scale observables in the presence of a long-wavelength fluctuation can be approximated
as evolving in a separate FRW universe. 
While the long-wavelength mode considered here must still be in the linear regime,
the small-scale observables do not since this equivalence is holds nonlinearly.   
As in the case of general relativity, by matching the expansion history of
the synchronous observers to the background, we can provide simple and accurate predictions for these observable
effects through small-scale cosmological simulations but now in any modified gravity theory that 
obeys this condition. 

  This equivalence leads to consistency
relations between changes in cosmological parameters and observable responses to
perturbations, for example through the  angle-averaged  squeezed $N$-point correlation functions or the
bias of dark matter halos. It would be interesting to work out the detailed statements of these relations for certain gravitational theories. Violation of these consistency relations indicate new physics beyond that encapsulated by 
the cosmological background.   We have also determined the scale at which these
violations could occur as a function of the metric perturbations themselves.

Even below the scale at which the separate universe ansatz fails, observables that
respond directly only to the local expansion history and not the  local curvature can be
accurately modeled by matching the former through ``fake" stress energy
components designed to mimic the effects of evolving curvature.  
These include the squeezed $N$-point correlation functions and the halo bias
in dynamical dark energy models as has been explicitly tested in simulations
 \cite{Hu:2016ssz,Chiang:2016vxa}.  
 
  In a modified gravity theory with screening mechanisms
 on small scales (see, e.g.~\cite{Joyce:2014kja}), the same principles should hold so long as the inhomogeneity of the
 long-wavelength mode does not enter directly through its spatial derivatives.   
We leave these considerations to  a future work.

\vfill
\acknowledgments{
We thank Hayato Motohashi and Marco Raveri  for useful discussions.  
WH was supported by U.S.~Dept.\ of Energy
contract DE-FG02-13ER41958,  NASA ATP NNX15AK22G. AJ was supported 
by NASA ATP grant NNX16AB27G and by the Robert R. McCormick Postdoctoral Fellowship. This work was further supported by 
the Kavli Institute for Cosmological Physics at the University of
Chicago through grants NSF PHY-0114422 and NSF PHY-0551142 and an endowment from 
the Kavli Foundation and its founder Fred Kavli.  WH thanks the Aspen Center for Physics, which is supported by National Science Foundation grant PHY-1066293, 
where part of this work was completed. }

\bibliography{SUmetric}

\end{document}